# Weak antilocalization and electron-electron interaction in coupled multiple-channel transport in a Bi$_2$Se$_3$ thin film

Yumei Jing[1], Shaoyun Huang[1,a)], Kai Zhang[2], Jinxiong Wu[2], Yunfan Guo[2], Hailin Peng[2], Zhongfan Liu[2], and H. Q. Xu[1,3,a)]

[1] Key Laboratory for the Physics and Chemistry of Nanodevices and Department of Electronics, Peking University, Beijing 100871, China

[2] Center for Nanochemistry, Beijing National Laboratory for Molecular Sciences (BNLMS), State Key Laboratory for Structural Chemistry of Unstable and Stable Species, College of Chemistry and Molecular Engineering, Peking University, Beijing 100871, China

[3] Division of Solid State Physics, Lund University, Box 118, S-221 00 Lund, Sweden

(December 9, 2015)

**Abstract**

Electron transport properties of a topological insulator Bi$_2$Se$_3$ thin film are studied in Hall-bar geometry. The film with a thickness of 10 nm is grown by van der Waals epitaxy on fluorophlogopite mica and Hall-bar devices are fabricated from the as-grown film directly on the mica substrate. Weak antilocalization and electron-electron interaction effects are observed and analyzed at low temperatures. The phase-coherence length extracted from the measured weak antilocalization characteristics shows a strong power-law increase with decreasing temperature and the transport in the film is shown to occur via coupled multiple (topological surface and bulk states) channels. The conductivity of the film shows a logarithmically decrease with decreasing temperature and thus the electron-electron interaction plays a dominant role in quantum corrections to the conductivity of the film at low temperatures.

PACS number(s): 73.20.Fz, 72.15.Rn, 73.25.+I, 73.50.-h

[a)] Authors to whom correspondence should be addressed. Electronic addresses: hqxu@pku.edu.cn and syhuang@pku.edu.cn.

Three-dimensional (3D) topological insulators (TIs) manifest themselves with gapless linear energy-dispersive metallic surface states of helical Dirac fermions in the insulating bulk band gaps.[1] The time-reversal symmetry protected topological surface states have consequently attracted considerable attention in recent years due to the fact that these states are of the fundamental interest in searching for novel physics phenomena, including elusive Majorana fermions in solid state, and have the potential applications in low energy-dissipative spintronics and quantum information processing.[2] The π-Berry phase associated with the helical Dirac fermion surface states gives rise to a quantum correction to the conductivity at low temperatures. The correction has the same forms of temperature and magnetic field dependences as the weak antilocalization (WAL) effect found in a two-dimensional (2D) electron system in the presence of strong spin-orbit interaction and is thus also termed as the WAL effect.[3-5] The WAL effect is suppressed when the time reversal symmetry is broken by applying a magnetic field. The effect can also be faded away as the phase coherence time $\tau_\varphi$ of electrons is decreased. At low temperatures, decoherence arises dominantly from electron-electron interaction (EEI) and thus $\tau_\varphi$ varies with temperature $T$ as $\propto T^{-p}$ with $p = 1$ for a 2D electron system.[6] The EEI also gives rise to an additional correction to the conductivity which has a similar form of temperature dependence as the WAL effect, but is competing with the WAL correction. Both the WAL effect and EEI effect have been observed in the transport studies of 3D TI thin films.[7] While, in many cases, the observed WAL correction to the conductivity agrees well with the theoretical prediction, the correction to the conductivity by the EEI effect observed in experiments could not always be accounted for by theory.[8]

In this paper, we report on a study of transport properties of a $Bi_2Se_3$ thin film grown by van der Waals epitaxy on fluorophlogopite mica. We analyze the quantum corrections to the conductivity induced by WAL and EEI at low temperatures in details and present a consistent description of EEI processes in the thin film. The film employed is of ~10 nm in thickness, which is thick enough to form decoupled helical band states on two surfaces and in the same time is thin enough for bulk transport to be of the 2D or quansi-2D nature. We show that the electron phase coherence length $l_\varphi$ in our thin film determined from the WAL effect is increased with decreasing temperature $T$ as $l_\varphi \sim T^{-0.72}$ and the transport occurs via coupled multiple (topological surface states and bulk states) channels in the film. We also show that the conductivity of the thin film is decreased logarithmically with decreasing temperature at $T <$ ~10 K and the EEI plays a dominant role in the quantum corrections to the conductivity at low temperatures. Our results are consistent with theory.

Among recently discovered three-dimensional TIs, confirmed by measurements using the surface-sensitive techniques of angle-resolved photoemission spectroscopy and scanning tunneling microscopy,[9,10] $Bi_2Se_3$ possesses a single Dirac cone in the surface-state band structure and a large bulk-state bandgap, and is hence considered as one of the most important TIs for the study of topology-protected surface states and for the applications using the helical surface Dirac fermions. So far, $Bi_2Se_3$ thin films have been prepared by mechanical cleavage, molecular beam epitaxy (MBE), and chemical vapor deposition methods. Our rhombohedral $Bi_2Se_3$ thin film, consisting of quintuple layers (QLs) stacked along the [0001] direction, is grown via van der Waals epitaxy on an atomically smooth, exfoliated fluorophlogopite mica sheet substrate in a vapor-phase deposition system.[11] It has been expected that a $Bi_2Se_3$ thin film grown by van der Waals epitaxy on fluorophlogopite mica is of high crystal quality and has excellent transport properties. In the van der Waals epitaxial growth for this work, the source material $Bi_2Se_3$ powder (Alfa Aesar, purity 99.999%) is placed in the center of a horizontally arranged one-inch-diameter quartz tube inside a twelve-inch tube furnace for thermal evaporation. A fluorophlogopite mica substrate is placed at a certain location inside the tube in order to set a desired deposition temperature. Before the vapor-phase deposition process, the tube is pumped to low pressure of ~50 mTorr and is flushed repeatedly with ultrapure argon to minimize oxygen contamination. Then, the source is heated to a temperature of 490 ºC and argon is used as a carrier gas to transport the source vapor to the mica substrate in the colder region of the furnace. The thickness of the $Bi_2Se_3$ thin film can be tuned with growth time and deposition temperature. The growth time of one minute at a pressure of 50 Torr, a carrier gas flow rate of 500 sccm and a deposition temperature of 390 ºC gives rise to a ~10-nm-thick $Bi_2Se_3$ crystalline thin film.

Figure 1(a) shows an optical microscope image of an as-grown $Bi_2Se_3$ thin film. During the epitaxial growth process, $Bi_2Se_3$ first nucleates at random locations on the mica substrate and then grows into triangular or hexagonal nanoplate islands. Eventually, these islands coalesce to form 2D nanosheet aggregates and finally a continuous film on the mica. Figure 1(b) shows an atomic force microscope (AFM) image of an as-grown $Bi_2Se_3$ film. Here, terraces with atomically smooth surfaces and steps of ~1 nm in height are clearly seen. The overall root-mean-square surface roughness of the thin film is measured and is found to be less than 1 nm, revealing an ultra-smooth surface. Compared with the TI thin films grown by MBE,[12] the terraces here are large in size with an area of up to a few $\mu m^2$. Figure 1(c) shows a typical Raman spectrum of the as-grown $Bi_2Se_3$ film, taken under an excitation using a 514-nm-wavelength laser. Three characteristic peaks centering at 71 cm$^{-1}$, 131 cm$^{-1}$, 173 cm$^{-1}$ are clearly visible in the measured Raman spectrum. These three vibration modes can be identified as of $A_{1g}^1, E_g^2$, and $A_{1g}^2$ ones, in agreement with the characteristic vibration modes of a few-quintuple-layer $Bi_2Se_3$ film on a mica substrate.[13] Transmission electron microscope (TEM, FEI Tecnai F30) characterization is carried out to further examine the crystalline quality of the $Bi_2Se_3$ film. Figure 1(d) shows a typical high-resolution TEM image of an as-grown $Bi_2Se_3$ film. Here, the expected hexagonal lattice fringes with a lattice spacing of 0.21 nm, consistent with interplanar spacing of the (11$\bar{2}$0) planes of rhombohedral $Bi_2Se_3$, can be found, suggesting that the film is of high crystalline quality.[14]

Hall-bar shaped devices, as schematically shown in the inset of Fig. 2(a), are fabricated from an as-grown, ~10-nm-thick, $Bi_2Se_3$ thin film on the growth mica substrate. Metal electrodes of Ti/Au (5/90 nm) are prepared using ultraviolet (UV) photolithography, electron-beam metal evaporation, and lift-off techniques. The Hall-bars with a core size of 50 μm (*L*) × 50 μm (*W*) are defined by a second UV photolithography and an argon-plasma etching process. On the mica substrate, 24 devices are fabricated and are arranged in two groups. The substrate is then cut into two pieces, named as Samples A and B, each containing 12 devices. Sample A is measured in a Quantum Design physical property measurement system (PPMS) and Sample B is measured in an Oxford dilution refrigerator to cover temperatures from 300 K down to 0.04 K. The magnetic-field dependences of the longitudinal sheet resistance $R_{xx}(B)$ and the transverse Hall-resistance $R_{xy}(B)$ are recorded using a lock-in setup with an excitation current of 300 nA at 17 Hz and the magnetoconductivity tensor is obtained from the measured values of $R_{xx}(B)$ and $R_{xy}(B)$. In the followings, we focus on the representative results of the measurements on two devices—one from Sample A (labeled as Device A) and one from Sample B (labeled as Device B).

Figure 2(a) shows the results of the measurements of Device A at 2 K at magnetic fields applied perpendicularly to the $Bi_2Se_3$ film. It is seen that the normalized longitudinal sheet resistance displays a sharp dip at zero magnetic field. This resistance dip is known as the WAL effect, derived from the π-Berry-phase topological delocalization.[4,5] It is also seen that the measured longitudinal sheet resistance $R_{xx}(B)$ shows a positive magnetoresistance over a large range of the applied magnetic fields. This positive magnetoresistance is in agreement with the results of the measurements carried out for $Bi_2Se_3$ thin films grown by MBE and hot wall CVD.[15,8] In our measurements, the line shape of the magnetoresistance $R_{xx}(B)$ at low magnetic fields (|*B*| < ~5 T) can be described by a power law function of $[R(B) - R(0)]/R(0) \propto |B|^\gamma$ with γ = 1.1. However, at high magnetic fields, the magnetoresistance $R_{xx}(B)$ shows a linear magnetic field dependence. Previously, such a linear magnetic field dependent behavior at high magnetic fields was observed in silver chalcogenides and was explained by Parish and Littlewood[16] using a network model accounting for macroscopic disorder and strong inhomogeneity in the materials. Recently, similar high-field, linear magnetic field dependent behaviors have been observed in MBE grown $Bi_2Se_3$ thin films by He *et al.*[17] and in

CVD grown Bi$_2$Se$_3$ thin films by Yan *et al.*[18] While Yan *et al.* have presented an evidence which supports the Parish-Littlewood model, He *et al.* have shown that the observed linear magnetic field dependent property of the magnetoresistance in thick Bi$_2$Se$_3$ films are associated with gapless energy spectrum of surface Dirac fermions and can thus be traced to the quantum origin of Abrikosov.[19] The measured Hall resistance $R_{xy}(B)$ displayed in Fig. 2(a) shows a good linear line shape at low fields. However, at high fields, a weak deviation from the linear magnetic field dependence can be seen in the measured $R_{xy}(B)$, which might indicate that multiple (topological surface states and bulk states) transport channels are present in our thin Bi$_2$Se$_3$ film.[20,21]

Figure 2(b) shows the temperature dependences of the sheet electron concentration and the electron mobility extracted for Device A at temperatures from 300 to 2 K. It is seen that the sheet carrier density decreases drastically with decreasing temperature from 100 K and then shows slow decreases at temperatures below ~10 K. The mobility is seen to increase with decreasing temperature and shows saturation at temperatures below ~10 K, indicating that phonon scattering in the thin film is frozen out when the temperature is lower than ~10 K. From Fig. 2(b), it can be seen that at $T = 2$ K, the sheet electron concentration is $n_s \sim 6.1\times10^{13}$ cm$^{-2}$ and the electron mobility $\mu$ is ~ 472 cm$^2$V$^{-1}$s$^{-1}$ in our Bi$_2$Se$_3$ thin film. The electron mean free path extracted for the film is then $l_e = 61$ nm. This value is smaller than the size of the Hall bars and the transport is thus in the diffusive regime.

Figure 3(a) shows the magnetoconductivity, $\Delta\sigma_{xx}(B) = \sigma_{xx}(B) - \sigma_{xx}(0)$, of Device B measured at the base temperature of 40 mK with tilted magnetic fields applied in the dilution refrigerator. It is seen that the magnetoconductivity peak is gradually broadened when the magnetic field is rotated from the perpendicular direction to the parallel direction (with the respect to the substrate). Figure 3(b) shows the measured magnetoconductivity against the perpendicular component of the applied magnetic field, i.e., $B\sin\theta$, where $\theta$ is the angle between the applied magnetic field and the film plane as indicated schematically in the inset of Fig. 3(b). The in-plane component of the magnetic field is perpendicular to the current direction. It is seen that the measured magnetoresistivity is solely dependent on the perpendicular component of the applied magnetic field. This result indicates that the transport in our thin film is dominantly of the 2D nature at this low magnetic field region. This is consistent with the fact that the thickness of the film is much smaller than the phase coherence length (see below) and the carrier mean free path in the film. We note that the magnetoconductivity curve is still seen to be peak like at $\theta = 0°$, which indicates that the magnetic field is not exactly in plane at $\theta = 0°$. To estimate the possibly misaligned angle $\Delta\theta$ with respect to $\theta = 0°$, we normalize the measured curve at $\theta = 0°$ against $B\sin(\theta + \Delta\theta)$ and find out that the misalignment $\Delta\theta$ is only about 2º.

Figure 4(a) shows the magnetoconductivity $\Delta\sigma_{xx}(B)$ of the devices measured in the magnetic field $B$ applied perpendicularly to the device surfaces at several representative temperatures between 0.04 K and 20 K. Here, both the results of the measurements of Device A at temperatures from 1.85 K to 20 K and the results of the measurements of Device B at temperatures from 0.04 K to 1.5 K are presented. It is seen that the magnetoconductivity peak at low fields is broadened with increasing temperature and eventually vanishes when the temperature becomes higher than 20 K. In the limit of strong spin-orbit interaction in a 2D system, i.e., $\tau_{so} \ll \tau_\varphi$ ($\tau_{so}$ is the spin-orbit scattering time), which is applicable for our devices, the quantum correction to the magnetoconductivity at low fields can be described by the simplified Hikami-Larkin-Nagaoka (HLN) formula as given by[3]

$$\Delta\sigma_{xx}(B) = \sigma_{xx}(B) - \sigma_{xx}(0)$$
$$= -\alpha \frac{e^2}{2\pi^2 \hbar} \left[ \psi\left(\frac{B_\varphi}{B} + \frac{1}{2}\right) - \ln\left(\frac{B_\varphi}{B}\right) \right], \quad (1)$$

where $\psi(x)$ is the digamma function, $B_\varphi = \hbar/(4el_\varphi^2)$ is the critical magnetic field characterized by the phase-coherence length $l_\varphi$, and $\alpha$ is the pre-factor that counts for the number of conduction channels ($\alpha = 1/2$ when only a single conduction channel is present in the transport). Our measured magnetoconductivity in the magnetic field range of $|B| < 0.15$ T can be well fitted by the simplified HLN formula, Eq. (1). Figure 4(b) shows the values of $l_\varphi$ and $\alpha$ extracted from fittings of the measurements at the temperatures between 0.04 K and 10 K. Here, the extracted values for Device A are marked by solid dots and the extracted values for Device B are marked by solid triangles. It is seen that the two sets of extracted values are in good agreement, implying the $Bi_2Se_3$ thin film is of high quality and uniform. It is also seen that $\alpha$ stays at a constant value of 0.6±0.01 for Device A and a constant value of 0.65±0.03 for Device B. The fact that $\alpha$ is closer to 1/2 rather than 1 could be due to the fact that the coupling between the surface states and the metallic bulk states in the thin film makes multiple transport channels (topological surface states and bulk states) not completely separable.[15] To check the applicability of the simplified HLN formula in our analyses, the WAL data measured at T = 1.85 K are also fitted by using the full expression of the HLN theory as described in Ref. [3], taking into account elastic scattering, dephasing scattering, and spin-orbit scattering processes. The fitting yields the phase-coherence length $l_\varphi \approx 634$ nm and the spin-orbit length $l_{so} \approx 9$ nm, agreeing well with the criterion of $\tau_{so} \ll \tau_\varphi$. Moreover, $l_\varphi$ happens to be the same as what we have extracted by using the simplified HLN formula, indicating that the simplified HLN formula is applicable and very accurate in the analyses of our measurements.

In Fig. 4(b), it is seen that the phase-coherence length $l_\varphi$ increases with decreasing temperature and starts to show saturation to a value of ~ 900 nm at temperatures below 600 mK. This saturation is a feature commonly found in dirty metals, semiconductors and ballistic quantum wires and dots.[22-24] The physical origin of the observed $l_\varphi$ saturation is still in debate and there is no generally accepted process which can satisfactorily explain all relevant experimental results. With regards to the measurements presented in this work, we would like to attribute the observed $l_\varphi$ saturation to electron heating and non-equilibrium effects that could arise from using the excitation current of 300 nA. This is because, as we have checked, the temperature at which $l_\varphi$ starts to show saturation is found to decrease to 400 mK when our measurements are carried out at the excitation current of 200 nA. In the temperature range of 1 to 10 K, phase breaking process is primarily due to EEI. In this temperature range, the measured phase-coherence length in our experiment can be described by a power law dependent function of $l_\varphi \sim T^{-\beta}$ with $\beta = 0.72$. This value of $\beta$ differs from the value of $\beta = 0.5$ predicted theoretically for EEI-induced phase breaking in a 2D electron system.[6] However, it is consistent with the results shown above that the transport occurs via coupled multiple (2D topological surface states and 2D bulk states) channels in our thin film.

Figure 5(a) shows the sheet resistance $R_{xx}$ of the $Bi_2Se_3$ thin film measured as a function of temperature in the absence of a magnetic field. As shown in the figure, $R_{xx}$ decreases monotonically with decreasing temperature from 300 to 10 K. This typical metallic behavior has often been observed in $Bi_2Se_3$ films.[25] At temperatures below 10 K, however, $R_{xx}$ turns to increase logarithmically. Figure 5(b) shows the conductivity $\sigma_{xx}(T)$ measured at various magnetic fields in this low temperature region. Importantly, the slope of $\sigma_{xx}(T)$ is seen to increase with increasing magnetic field and is approximately unchanged when $B > 1$ T. The considerably decreased conductivity with decreasing temperature at finite magnetic fields reflects an insulating ground state at the low temperature limit. The coupling between the bottom and the top topological surface states in an ultrathin film may open a sizeable energy gap at Dirac cone point and could lead to insulating ground state at low temperatures.[26] The presence of the gapless topological surface states, however, is indicated consistently by a number of ARPES measurements within their energy resolutions in $Bi_2Se_3$ thin films with a thickness similar to our sample (10 nm).[27] Therefore, this insulating ground state is most likely induced by the EEI effect in our thin film at low temperatures.

In the absence of a magnetic field, both the weak localization (WL) and the EEI in a diffusive metallic system yield logarithmical temperature dependent behaviors of the conductivity $\sigma_{xx}$. However, the magnetotransport measurements have already shown a magnetoresistance dip at zero magnetic field and thus the WAL effect rather than WL effect is present in our $Bi_2Se_3$ film in the absence of a magnetic field. The WAL effect makes $\sigma_{xx}$ increase with decreasing temperature and the resulting quantum correction from the π-Berry-phase delocalization to $\sigma_{xx}$ at different temperatures

can be described as[28]

$$\Delta\sigma_{xx}^{WAL}(T) = -\frac{\alpha p e^2}{2\pi^2 \hbar}\ln\left(\frac{T}{T_L}\right), \quad (2)$$

where $T_L$ is a characteristic temperature at which the WAL correction vanishes, $\hbar$ is the reduced Plank constant, and $p$ is the temperature exponent in the inelastic scattering time $\tau_i \propto T^{-p}$ ($p$ is related to $\beta$ as $p=2\beta$). On the other hand, the EEI modifies $\sigma_{xx}$ oppositely as[29]

$$\Delta\sigma_{xx}^{EEI}(T) = \frac{e^2}{2\pi^2 \hbar}\left(1-\frac{3}{4}F\right)\ln\left(\frac{T}{T_E}\right), \quad (3)$$

where $F$ is the electron screening factor whose value lies between 0 and 1 theoretically and $T_E$ is a characteristic temperature at which the EEI correction is suppressed. Therefore, the logarithmical decrease in $\sigma_{xx}$ at temperatures below 10 K could derive from the competitive WAL and EEI corrections. The screening factors $F$ can be determined by fitting the data of Fig. 5(b) with Eq. (3) at a magnetic field $B \gtrsim 1$ T, which is much larger than the critical magnetic field beyond which the WAL is suppressed. Here the critical field can be estimated from $B_\varphi \approx \hbar/4el_\varphi^2$, which is about 0.0073 T at $T = 10$ K [using $l_\varphi \approx 150$ nm as shown in Fig. 4(b)] and gets smaller at lower temperatures in our film. The screening factor $F$ is found to lie between 0 and 0.15, well close to 0, implying that strong EEI is present in our thin Bi$_2$Se$_3$ film. At $B = 0$ T, the combination of the WAL and EEI effects allows us to extract the fitting factor $\overline{F} = 1 - \frac{3}{4}F - \alpha p = 0.042$. This result shows that in the absence of a magnetic field, the EEI plays the same important role in the quantum correction to the conductivity of the thin film as the WAL. We emphasize, however, again that in the presence of a sufficiently large magnetic field, the quantum correction to the conductivity is dominantly due to the EEI in the thin film. By using an averaged value of $F = 0.08$, we can obtain $\alpha p = 0.9$, which is consistent with the value of $2\alpha\beta = 0.87$, determined from the phase-coherence length measurements.

In conclusion, we have studied the electric transport properties of a Bi$_2$Se$_3$ thin film grown by vapor-phase deposition in Hall-bar geometry. The film is characterized by magnetotransport measurements and it is shown that the film possesses an n-type sheet electron concentration of $n_s \sim 6.1\times10^{13}$ cm$^{-2}$ and an electron mobility of $\mu \sim 472$ cm$^2$V$^{-1}$s$^{-1}$ at $T = 2$ K. At high temperatures (10 to 300 K), typical metallic behavior is observed in the thin film, namely, the resistivity is decreased with decreasing temperature. At low temperatures (< 10 K), the WAL effect and EEI effect are observed in the transport measurements. The WAL effect is analyzed based on the HLN theory and the phase coherence length of electrons in the thin film is extracted. It is shown that the phase coherence length is scaled with temperature as $l_\varphi \sim T^{-\beta}$ with $\beta = 0.72$ at 0.04 to 10 K and the transport in our thin film occurs via coupled multiple (2D topological surface states and 2D bulk states) channels. The effect of EEI is studied by measuring the conductivity of the thin film as a function of temperature. It is found that at temperatures below ~10 K, the conductivity of the thin film shows a logarithmical decrease with decreasing temperature at both zero magnetic field and high magnetic fields at which the WAL is suppressed, revealing the presence of a significant effect of EEI in the thin film. Further experiment with $Bi_2Se_3$ thin films of tunable Fermi level is needed in order to switch on and off the coupling between the surface states and bulk states and, thus, a detailed study of the quantum corrections originating purely from the surface state transport in the thin films can be carried out.[30]

We thank Prof. Alex Hamilton at University of New South Wales for helpful discussion. This work was supported by the National Basic Research Program of China (Nos. 2012CB932700 and 2012CB932703), the National Natural Science Foundation of China (Nos. 91221202, 91421303, 11274021, and 61321001), and the Specialized Research Fund for the Doctoral Program of Higher Education of China (No. 20120001120127). HQX also acknowledges financial support from the Swedish Research Council (VR).

**References**

[1] M. Z. Hasan and C. L. Kane, Rev. Mod. Phys. **82**, 3045 (2010).

[2] X. L. Qi and S. C. Zhang, Rev. Mod. Phys. **83**, 1057 (2011).

[3] S. Hikami, A. I. Larkin, and Y. Nagaoka, Prog. Theor. Phys. **63**, 707 (1980).

[4] P. Ghaemi, R. S. K. Mong, and J. E. Moore, Phys. Rev. Lett. **105**, 166603 (2010).

[5] G. Tkachov and E. M. Hankiewicz, Phys. Rev. B **84**, 035444 (2011)

[6] B. L. Altshuler, A. G. Aronov, and D. E. Khmelnitsky, J. Phys. C **15**, 7367 (1982).

[7] J. H. Bardarson, and J. E. Moore, Rep. Prog. Phys. **76**, 056501 (2013).

[8] Y. Takagaki, B. Jenichen, U. Jahn, M. Ramsteiner, and K. J. Friedland, Phys. Rev. B **85**, 115314 (2012).

[9] Y. Xia, D. Qian, D. Hsieh, L. Wray, A. Pal, H. Lin, A. Bansil, D. Grauer, Y. S. Hor, R. J. Cava, and M. Z. Hasan, Nat. Phys. **5**, 398 (2009).

[10] P. Cheng, C. L. Song, T. Zhang, Y. Y. Zhang, Y. L. Wang, J. F. Jia, J. Wang, Y. Y. Wang, B. F. Zhu, X. Chen, X. C. Ma, K. He, L. L. Wang, X. Dai, Z. Fang, X. C. Xie, X. L. Qi, C. X. Liu, S. C. Zhang, and Q. K. Xue, Phys. Rev. Lett. **105**, 076801(2010).

[11] H. Li, J. Cao, W. S. Zheng, Y. L. Chen, D. Wu, W. H. Dang, K. Wang, H. L. Peng, and Z. F. Liu, J. Am. Chem. Soc. **134**, 6132 (2012).

[12] X. F. Kou, L. He, F. X. Xiu, M. R. Lang, Z. M. Liao, Y. Wang, A. V. Fedorov, X. X. Yu, J. S. Tang, G. Huang, X. W. Jiang, J. F. Zhu, J. Zou, and K. L. Wang, Appl. Phys. Lett. **98**, 242102 (2011).

[13] W. Richter, H. Kohler, and C. R. Becker, Phys. Status Solidi B **84**, 619 (1977).

[14] H. L. Peng, W. H. Dang, J. Cao, Y. L. Chen, W. Wu, W. S. Zheng, H. Li, Z. X. Shen, and Z. F. Liu, Nat. Chem. **4**, 281 (2012).

[15] Y. S. Kim, M. Brahlek, N. Bansal, E. Edrey, G. A. Kapilevich, K. Iida, M. Tanimura, Y. Horibe, S. W. Cheong, and S. Oh, Phys. Rev. B **84**, 073109 (2011).

[16] M. M. Parish and P. B. Littlewood, Nature **426**, 162 (2003).

[17] H. T. He, B. K. Li, H. C. Liu, X. Guo, Z. Y. Wang, M. H. Xie, and J. N. Wang, Appl. Phys. Lett. **100**, 032105 (2012).

[18] Y. Yan, L. X. Wang, D. P. Yu, and Z. M. Liao, Appl. Phys. Lett. **103**, 033106 (2013).

[19] A. A. Abrikosov, Phys. Rev. B **58**, 2788 (1998).

[20] A. A. Taskin, S. Sasaki, K. Segawa, and Y. Ando, Adv. Mater. **24**, 5581 (2012).

[21] L. He, F. Xiu, X. Yu, M. Teague, W. Jiang, Y. Fan, X. Kou, M. Lang, Y. Wang, G. Huang, N. C. Yeh, and K. L. Wang, Nano Lett. **12**, 1486 (2012).

[22] J. J. Lin and J. P. Bird, J. Phys: Condens. Matter **14**, R501 (2002).

[23] H. Linke, P. Omling, H. Q. Xu, and P. E. Lindelof, Superlattices Microstruct. **20**, 441 (1996).

[24]H. Linke, P. Omling, H. Q. Xu, and P. E. Lindelof, Phys. Rev. B **55**, 4061 (1997).

[25]J. J. Cha, J. R. Williams, D. S. Kong, S. Meister, H. L. Peng, A. J. Bestwick, P. Gallagher, D. Goldhaber-Gordon, and Y. Cui, Nano Lett. **10**, 1076 (2010).

[26]J. Linder, T. Yokoyama, and A. Sudbø, Phys. Rev. B **80**, 205401 (2009).

[27]Y. Zhang, K. He, C. Z. Chang, C. L. Song, L. L. Wang, X. Chen, J. F. Jia, Z. Fang, X. Dai, W. Y. Shan, S. Q. Shen, Q. Niu, X. L. Qi, S. C. Zhang, X. C. Ma, and Q. K. Xue, Nat. Phys. **6**, 584 (2010).

[28]E. Abrahams, P. W. Anderson, D. C. Licciardello, and T. V. Ramakrishnan, Phys. Rev. Lett. **42**, 673 (1979).

[29]B. L. Altshuler, A. G. Aronov, and P. A. Lee, Phys. Rev. Lett. **44**, 1288 (1980).

[30]F. Yang, A. A. Taskin, S. Sasaki, K. segawa, Y. Ohno, K. Matsumoto and Y. Ando, Appl. Phys. Lett. **104**, 161614 (2014).

**Figure captions**

Fig. 1 (Color online) (a) Optical microscope image of a $Bi_2Se_3$ thin film grown by van de Waals epitaxy on mica. Here, a large-sized continuous film with triangular and hexagonal nanoplate islands on top is shown. (b) Atomic force microscope image of the film. Here, terraces with an atomic layer step height are observable. (c) Typical Raman spectrum taken with excitation by a 514-nm-wavelength laser. Here, three characteristic peaks centering at 71 cm$^{-1}$, 131 cm$^{-1}$, and 173 cm$^{-1}$ are clearly seen. (d) Typical high-resolution TEM image of the film. In the image, the hexagonal lattice fringes with a lattice spacing of 0.21 nm are seen.

Fig. 2 (Color online) (a) Longitudinal magnetoresistance and Hall resistance measured against the magnetic field applied perpendicularly to a $Bi_2Se_3$ thin film at 2 K. The inset shows schematics of the Hall-bar device and the measurement setup. (b) Measured electron mobility and concentration plotted against temperature. The dark and red curves are guides to the eyes.

Fig. 3 (Color online) (a) Magnetoconductivity measured as a function of the applied magnetic field with different orientations. (b) Magnetoconductivity plotted against the perpendicular component of the applied magnetic field. The inset in (b) shows schematically the orientation of the applied magnetic field.

Fig. 4 (Color online) (a) Magnetoconductivity measured in a low magnetic-field region at different temperatures. The results at temperatures from 1.85 to 20 K are obtained in a Quantum Design PPMS system (Device A) and the results at temperatures from 0.04 to 1.5 K are obtained in an Oxford dilution refrigerator (Device B). (b) Extracted number-of-channel parameters $\alpha$ and electron phase-coherence length $l_\varphi$ from the measurements shown in (a). The gray curve is guide to the eyes.

Fig. 5 (Color online) (a) Sheet resistance $R_{xx}$ measured as a function of temperature at $B$=0 T. (b) Sheet conductivity $\sigma_{xx}$ measured as a function of temperature at different magnetic fields. The straight lines are the results of least-square fits to the measurement data.

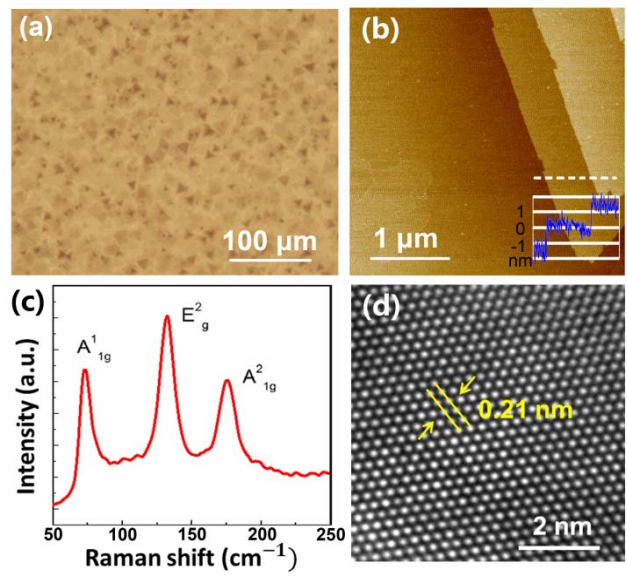

Fig. 1 by *Yumei Jing et al.*

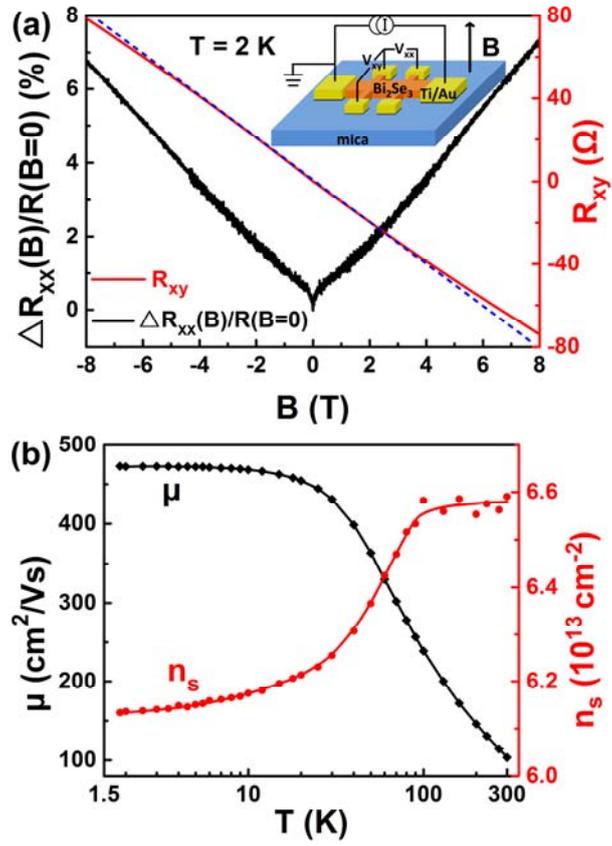

Fig. 2 by *Yumei Jing et al*

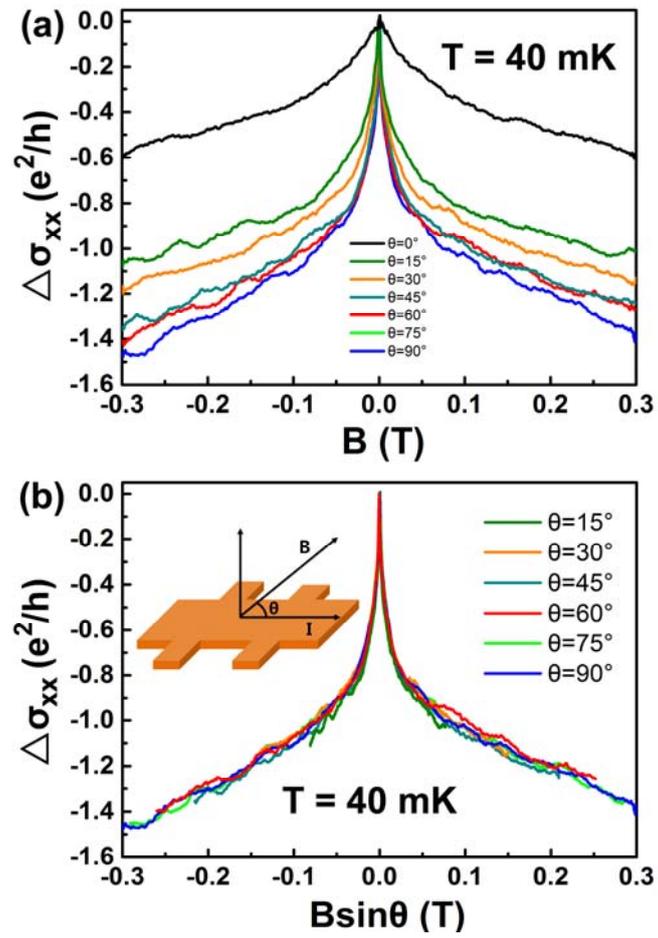

Fig. 3 by *Yumei Jing et al.*

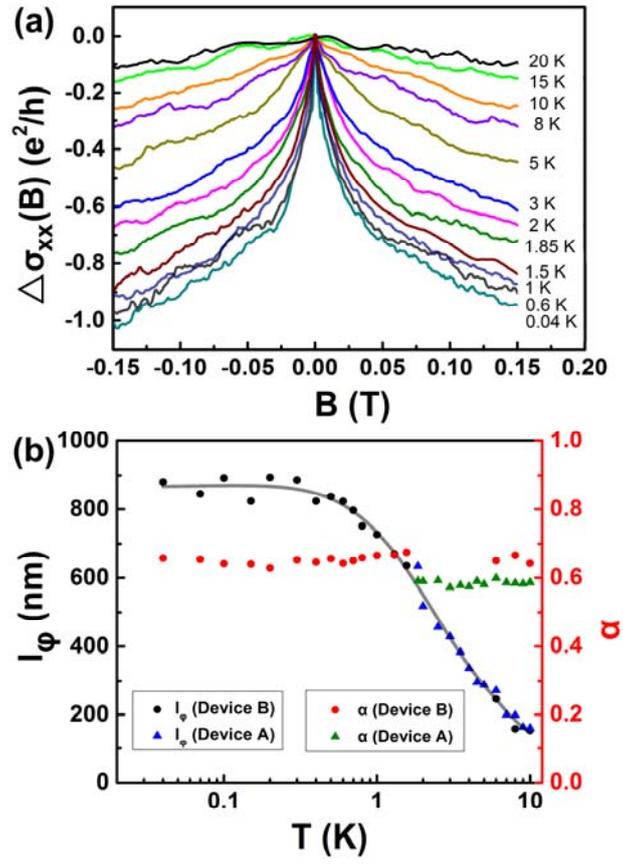

Fig. 4 by *Yumei Jing et al.*

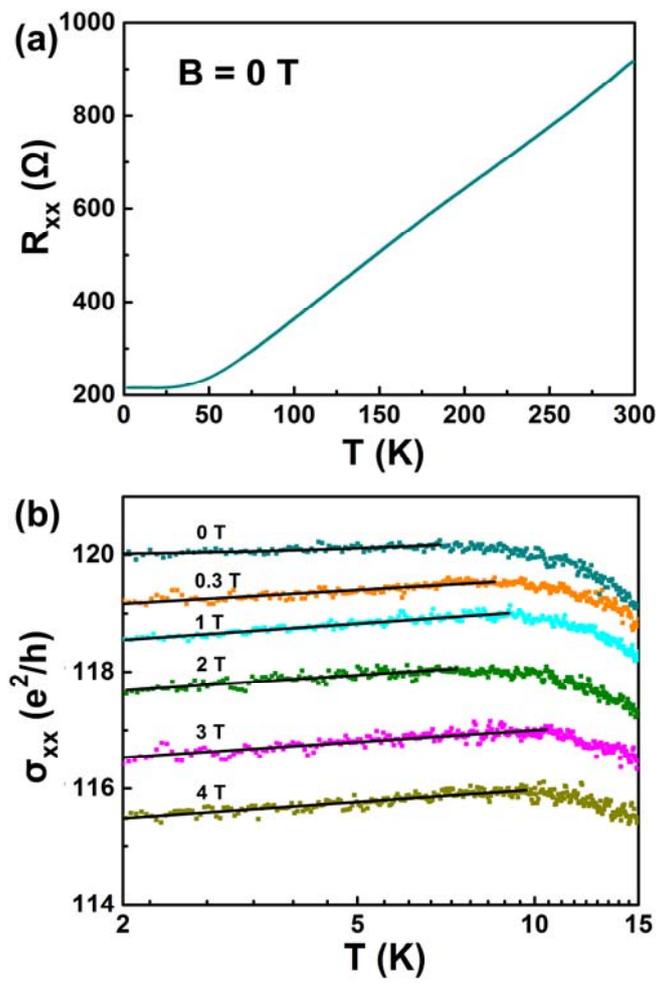

Fig. 5 by *Yumei Jing et al*